\documentclass[12pt,a4paper,notitlepage, reqno]{amsart}
\usepackage[utf8]{inputenc}
\usepackage[english]{babel}

\usepackage{bm,bbm}
\usepackage{amsbsy,amsmath,amsthm, amssymb,amsfonts}
\usepackage{graphicx}
\usepackage{subfigure}
\usepackage{verbatim}
\usepackage{accents}
\usepackage{stmaryrd}
\usepackage[foot]{amsaddr}
\usepackage{ifthen}
\usepackage{adjustbox}
\usepackage{float}
\usepackage{varwidth}

\theoremstyle{plain}
\newtheorem{theorem}{Theorem}[section]

\newtheorem{proposition}[theorem]{Proposition}

\theoremstyle{definition}

\theoremstyle{remark}

\usepackage[	pdfauthor={},%
			pdftitle={},%
			pdftex]{hyperref}
\hypersetup{	colorlinks,%
			citecolor=black,%
			filecolor=black,%
			linkcolor=black,%
			urlcolor=black%
			}

\usepackage[hmargin=3cm,vmargin={2.5cm,2cm},includefoot]{geometry}

\newcommand{\ddiamond}{\spadesuit}

\newcommand{\Z}{\mathbb{Z}}

\newcommand{\half}{\mathbb{H}}

\newcommand{\ii}{\mathrm{i}}

\renewcommand{\P}{\mathbb{P}}

\newcommand{\E}{\mathbb{E}}

\newcommand{\ind}{\mathbbm{1}}

\newcommand{\F}{\mathcal{F}}

\newcommand{\de}{\mathrm{d}}

\newcommand{\ffrac}[2]{\left(\frac{#1}{#2}\right)}

\newcommand{\eps}{\varepsilon}

\newcommand{\ggr}{G}

\newcommand{\lattice}{\mathbb{L}}
\newcommand{\sqlattice}{\lattice^\bullet}
\newcommand{\dsqlattice}{\lattice^\circ}
\newcommand{\dmdlattice}{\lattice^\diamond}
\newcommand{\ddmdlattice}{\lattice^\ddiamond}

\newcommand{\domain}{\Omega}

\newcommand{\intarcp}[4]{(#1 \frown #2, #3 \frown #4)}
\newcommand{\extarcp}[4]{(#1 \smile #2, #3 \smile #4)}

\newcommand{\sle}[1]{SLE$(#1)${}}
\newcommand{\slevar}[1]{SLE$[#1]${}}
\newcommand{\varsle}[1]{\slevar{#1}}

\begin{document}
\title{%
  Configurations of FK Ising interfaces and hypergeometric SLE}
%
%
%
\author{Antti Kemppainen$^1$}
\address{$^1$Department of Mathematics and Statistics\\
         P.O. Box 68\\
         FIN-00014 University of Helsinki\\
         Finland}
\author{Stanislav Smirnov$^2$}
\address{$^2$Section de math\'ematiques,
         Universit\'e de Gen\`eve,
         2-4, rue du Li\`evre, c.p. 64,
         1211 Gen\`eve 4, Switzerland, \textnormal{and} 
Chebyshev Laboratory, St.~Petersburg State University, Russia, \textnormal{and}  
Skolkovo Institute of Science and Technology, Russia}
\email{Antti.H.Kemppainen@helsinki.fi, Stanislav.Smirnov@unige.ch}
\date{\today}
\begin{abstract}
In this paper,
we show that the interfaces
in FK Ising model in any domain with 4 marked boundary points and
wired--free--wired--free boundary conditions conditioned
on a specific internal arc configuration of interfaces
converge in the scaling limit to hypergeometric SLE (hSLE).
The arc configuration consists of a pair of interfaces and the scaling limit of their joint law
can be described by an algorithm to sample the pair from a hSLE curve and a chordal SLE
(in a random domain defined by the hSLE).
\end{abstract}
\maketitle

\section{Introduction}

In the seminal paper \cite{Schramm:2000th}, Oded Schramm introduced SLE as a one-parameter family of 
conformally invariant random fractal curves,
and showed that those are the only possible conformally invariant scaling limits of 
the interfaces  in the lattice models at criticality.
The SLEs are dynamically grown, by running the Loewner evolution with a random driving term.
The original definition was formulated in two setups: 
chordal (curves between two boundary points) and radial (curves between a boundary and
an interior point), which both have trivial conformal modulus,
and thus their Loewner driving term is given by a Brownian motion without a drift. 
Soon afterwards Lawler, Schramm and Werner introduced a generalization \cite{LawlerEtal:2003}
for domains with several marked points
and the driving process drift having a very particular and elegant dependence on their conformal modules. 
While including several fundamental cases, this process does not cover all
the important situations, and it was quickly realized that one should also look at
more general SLEs, weighted by partition functions and having more complicated drifts 
\cite{Bauer:2005ki,Dubedat:2005ig,2013arXiv1309.5302I,Lawler:2007tb,Lawler:2009dv,Zhan:2008vd}.

In this paper we are concerned with a particular case of SLEs in a domain
with 4 marked boundary points connected in pairs by two non-intersecting SLE curves. 
Such arrangement corresponds to the wired-free-wired-free boundary conditions in the underlying FK model.
The marked boundary points can be connected in two ways, and conditioning on one of those 
we obtain the hypergeometric SLE, cf. \cite{Qian:2016wx,Wu:2016uz}.

\subsection{FK Ising model on \texorpdfstring{$\Z^2$}{Z2}}

Let $\sqlattice$ and $\dsqlattice$ be the even and odd sublattices of the square lattice $\Z^2$, 
respectively, that is,
the sum of the $x$ and $y$ coordinates is even or odd on $\sqlattice$ and $\dsqlattice$, respectively.
The lattices $\sqlattice$ and $\dsqlattice$ are both square lattices  with a lattice mesh $\sqrt{2}$.
The medial lattice $\dmdlattice$ is formed by the midpoints of edges of $\sqlattice$
(or equivalently of $\dsqlattice$) which then are connected with edges by going around
each face of $\sqlattice$. The graph $\dmdlattice$ is also a square lattice. The modified
medial lattice $\ddmdlattice$ is the square--octagon lattice which we get by replacing all
vertices of $\dmdlattice$ by a small square. See the introduction of 
\cite{Kemppainen:2015vu} for more information.

We call the octagons white or black, if their centers are in $\sqlattice$ and $\dsqlattice$, respectively.
Those faces of $\ddmdlattice$ that are squares are called small squares.

\begin{figure}[tbh]
\centering
\subfigure[A discrete domain with four marked boundary points.
Marked points are the cusp points in the picture. 
The boundary of the discrete domain consists of four ``admissible paths'' 
(boundaries of chains of octagons and small squares alternating) on $\ddmdlattice$.] 
{
	\includegraphics[scale=.275]
{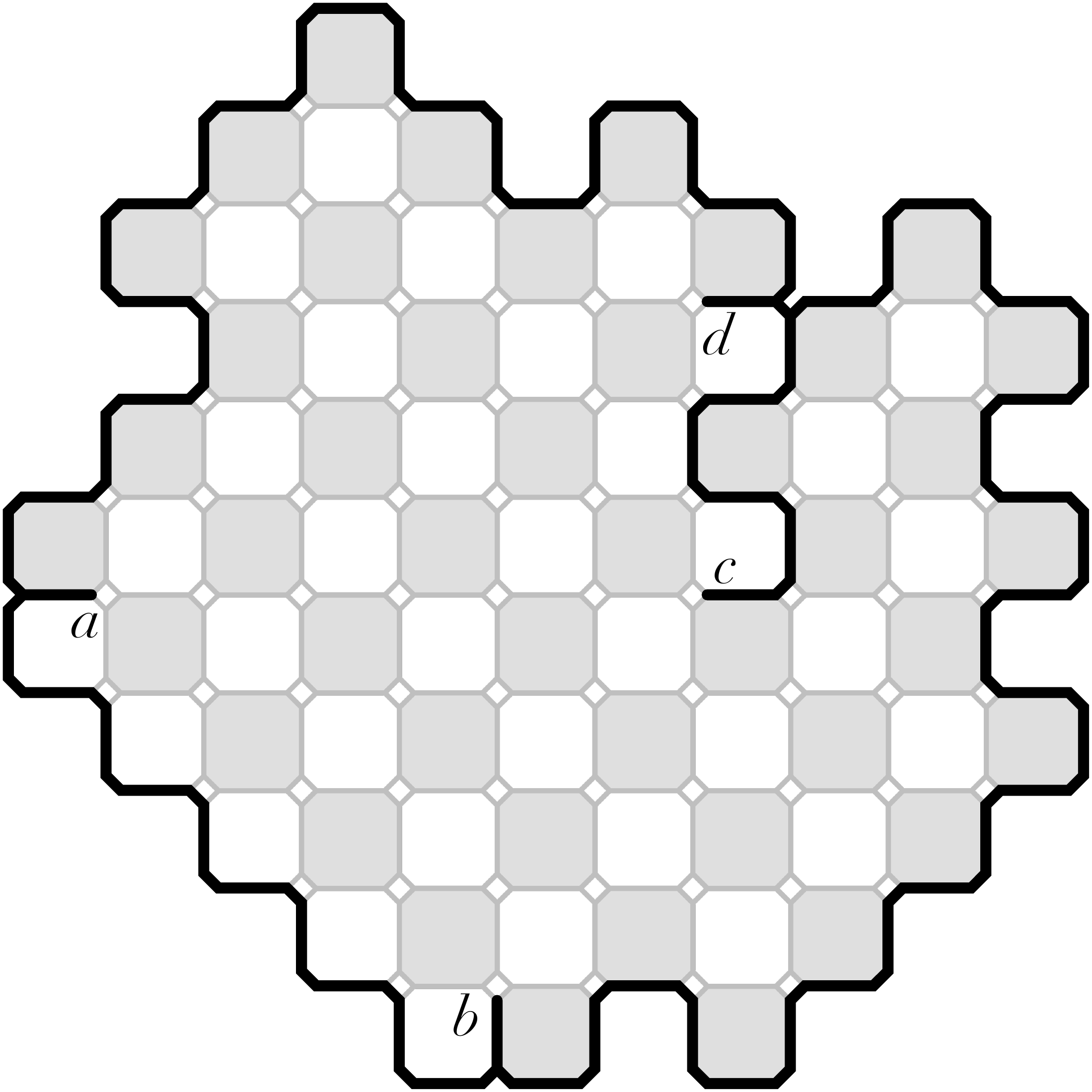}
} 
\hspace{0.5cm}
\subfigure[In the 4 marked boundary points setting,
a configuration of (the loop representation) of the FK Ising model
consists of two interfaces and a number of loops. 
If $a,b,c,d$ are as on the figure (a),
then the present configuration belongs to the internal arc configuration
event ${\intarcp{a}{d}{c}{b}}$.]
{
	\includegraphics[scale=.275]
{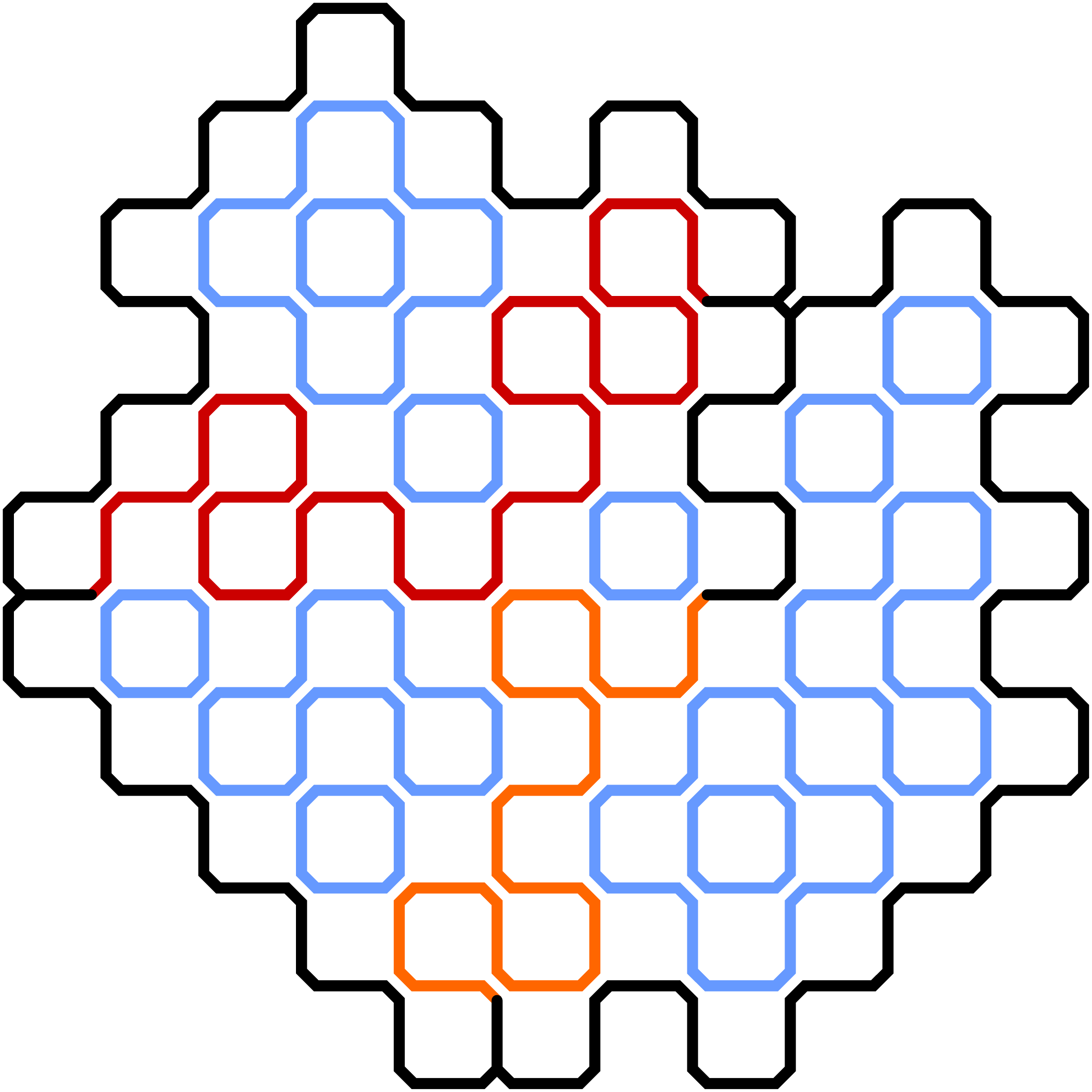}
}
\caption{Discrete domain with 4 marked boundary points and a loop configuration on it.}
\label{fig: graphs}
\end{figure}

Consider a domain $\domain$ whose boundary is the outer boundary of a simple closed  chain of 
faces where the octagons and the small squares are alternating. 
We assume that the ``boundary conditions'' change at 4 ``marked'' points, that is,
the chain consist of exactly 4 open monochromatic chains of octagons and small squares. 
See Figure~\ref{fig: graphs}. The marked points are denoted by $a,b,c,d$ in general and we can
assume they are the joint vertices of a black octagon, a white octagon and a small square in the chain
and thus on the boundary of the domain. Assume that $a,b,c,d$ are counterclockwise ordered
on the boundary and that the octagons next to the boundary arc $[ab]$ are white.
Then necessarily the octagons next to $[cd]$ are white as well and
the ones next to $[bc]$ or $[da]$ are black.

Let $\ggr^\ddiamond \subset \ddmdlattice$  be the graph which contains the vertices
$(V(\ddmdlattice) \cap \domain) \cup \{a,b,c,d\}$ and all the edges contained in $\domain$.
Let us consider a loop configuration which in the present case contains 2 open paths
that both connect  $\{a,c\}$ to $\{b,d\}$ and a number of closed loops.
We assume that the configuration is dense on $\ggr^\ddiamond$, in the sense
that it covers all the vertices, and that the paths are simple and mutually disjoint.
See Figure~\ref{fig: graphs}.

Define a probability measure on the dense simple loop configurations of $\ggr^\ddiamond$
by requiring that the probability of a
loop configuration is proportional to 
\begin{equation} \label{eq: def fki loop weight}
\sqrt{2}^{(\# \textnormal{ of loops})}.
\end{equation}
Notice that the number of open paths doesn't enter this formula, since there are
always 2 such paths. The model is called the loop representation of
the critical FK Ising model (Fortuin--Kasteleyn random cluster model with a parameter value that
corresponds to the Ising model).

\subsubsection{The motivation of the FK random cluster model}

The (spin) Ising model is a model for ferromagnetic substance. The Ising model configuration
is a field of $\pm 1$ random variables, one on each vertex of a graph, and their probability
law is given by the Boltzmann distribution of an energy functional with a nearest neighbor interaction.
A parameter $\beta$ determines the strength of the interaction.
The Fortuin--Kasteleyn random cluster model (FK model) 
is a percolation-type model with two parameter $p \in [0,1]$ and $q \geq 0$.
The FK model configuration is a random subset of the set of edges of a graph.
These edges are called open and the edge in the complement are called closed.
A connected component (of vertices) in that random graph is called a cluster.
In the FK model, the probability of the configuration is proportional to
$q^{(\# \textnormal{ of clusters})}\, p^{(\# \textnormal{ of open edges})}
   \, (1-p)^{(\# \textnormal{ of closed edges})}$.

The FK Ising model is a particular case $q=2$ of the FK model. The spin Ising model and the FK Ising model
are connected by the Edwards--Sokal coupling, that is, there exists a random field on the vertices and edges
such that the marginal distribution of the random field on the vertices is the spin Ising model
and the marginal distribution of the random field on the edges is the FK Ising model.
For instance, spin correlations can be expressed in terms of connection probabilities using this coupling.

We consider only the case $q=2$ with the critical parameter $p = \frac{\sqrt{2}}{\sqrt{2}+1}$
in this article.
Also we consider only the square lattice $\Z^2$, although, we could relax that assumption.

The loop representation of the random cluster configuration is a dense set of loops
such that no loop intersects any open or dual-open edges (the loops are the boundaries of primal
and dual clusters).
The choice of critical parameter for the FK Ising random cluster model leads to
the weight \eqref{eq: def fki loop weight} for the loop representation.

\subsection{Setting and notation for the scaling limit}

\subsubsection{Discrete setting, conditional measure and the scaling limit}

For some $\delta>0$, $(\domain_\delta,a_\delta,b_\delta,c_\delta,d_\delta)$ 
be a simply  connected discrete domain with four marked boundary points and lattice mesh $\delta>0$,
that is, the boundary of $\domain_\delta$ is a path on $\delta \ddmdlattice$ with
properties given above.
We assume that the boundary arcs $\alpha_1 = [a_\delta b_\delta]$, $\alpha_2 = [b_\delta c_\delta]$,
$\alpha_3 = [c_\delta,d_\delta]$
and $\alpha_4 = [d_\delta a_\delta]$ are simple lattice paths on the modified medial lattice $\delta \ddmdlattice$
such that the first and last edges are edges between two octagons and 
$\alpha_1, \alpha_2^\leftarrow, \alpha_3, \alpha_4^\leftarrow$, where $\alpha^\leftarrow$ denotes the reversal of $\alpha$,
have white octagons and small squares to their left and black octagons and white squares on their right.

Let $G_\delta$ be the graph on $\delta \sqlattice$ corresponding to $\domain_\delta$
and consider the FK model with wired boundary conditions on $[b_\delta c_\delta]$ and
$[d_\delta a_\delta]$.
Define also an enhanced graph
$\hat G_\delta$ where we add the external arc pattern $\extarcp{a_\delta}{b_\delta}{c_\delta}{d_\delta}$ in the sense that
the wired arcs are counted to be in the same component and in the weight~\eqref{eq: def fki loop weight}, if the interface
starting at $a_\delta$ ends at $b_\delta$, then it is counted as a closed loop.

There are two interfaces $\gamma$ and $\gamma^*$ starting at $a_\delta$ and $c_\delta$ respectively.
Denote by $\P_\delta$ the probability law of $\gamma$ and by $\P_\delta^+$ the measure $\P_\delta$
conditional to the fact that $\gamma$ ends to $c_\delta$.

The scaling limits $\P= \lim_{\delta \to 0} \P_\delta$ and $\P^+ = \lim_{\delta \to 0} \P_\delta^+$ are considered below.

\subsubsection{Conformal transformation to the upper half-plane}

It is useful to describe the probability laws in the upper-half plane (or in another fixed reference domain).
We apply a conformal transformation such that the points $a_\delta,b_\delta,c_\delta,d_\delta$
are mapped to points $U_0^\delta,V_0^\delta,W_0^\delta,\infty$ respectively. Then
$U_0^\delta < V_0^\delta < W_0^\delta$. As $\delta \to 0 $, these points tend
to some points $U_0 < V_0 < W_0$.

We will consider simple curves starting at $U_0$ as Loewner evolutions. 
In particular, we assume that they are parametrized by the half-plane capacity.
The driving process is denoted by $U_t$ and three other marked points are $V_t,W_t$ and $\infty$.
In particular, $V_t$ and $W_t$ satisfy the Loewner equation driven by $U_t$.
Then also $U_t < V_t < W_t$.
Auxiliary processes are defined by setting
\begin{equation*}
X_t = V_t - U_t , \qquad
Y_t = W_t - V_t  .
\end{equation*}

\subsection{The hypergeometric \texorpdfstring{\sle{\frac{16}{3}}}{SLE(16/3)}}

The hypergeometric SLE (see \cite{Wu:2016uz} and such processes also appear in 
\cite{Dubedat:2005ig,Qian:2016wx,Zhan:2008vl}. Abbreviated to hSLE.) with parameter value $\kappa= \frac{16}{3}$ is defined by
giving the driving process
\begin{align} \label{eq: hsle driving}
\de U_t & = \frac{4}{\sqrt{3}} \de  B_t + \left( -\frac{2}{X_t} + \frac{2}{X_t+Y_t} 
   - \frac{4}{3} \frac{Y_t\left( -1 + \sqrt{1+\frac{X_t}{Y_t}}\right)}{X_t(X_t+Y_t)}
  \right) \de t  .
\end{align}
Note that the third term inside the brackets
is equal to $-\frac{16}{3} \frac{F'(z)}{F(z)} \frac{1-z}{s}$ evaluated at $z=\frac{X_t}{X_t+Y_t}$ and $s=X_t+Y_t$,
where $F(z)$ is the hypergeometric function $\vphantom{F}_2 F_1(\frac{3}{4}, \frac{1}{4}; \frac{3}{2};z)$.

\subsection{The main result}

By the results of \cite{Kemppainen:2015vu}, the scaling limit  $\P= \lim_{\delta \to 0} \P_\delta$ 
is equal to a certain \varsle{\frac{16}{3},Z} process, that is, an \sle{\frac{16}{3}}
process with a partition function $Z$. The topology of the convergence
is given by the weak convergence of probability measures on the metric space of 
continuous functions. We use that result to prove the following theorem.

\begin{theorem}
The sequence $\P_\delta^+$ converges, in the same topology as above, to $\P^+$ which
is the law of a hypergeometric \sle{\frac{16}{3}}.
\end{theorem}

See also Section~\ref{ssec: joint law} below for description of the scaling limit of
the joint law of the pair $(\gamma,\gamma^*)$.

\section{Scaling limit of FK Ising model interface as hyperbolic SLE}

\subsection{Discrete martingale observable}

In the random cluster model we take
the boundary conditions which are free--wired--free--wired and they change
across the edges  corresponding to
$a_\delta,b_\delta,c_\delta,d_\delta$.
There are interfaces starting and ending to these points.
Due to topological (as well as parity) reasons, the interface starting
at $a_\delta$ has to end at $b_\delta$ or $d_\delta$. We denote
these two mutually exclusive events as
$\intarcp{a_\delta}{b_\delta}{c_\delta}{d_\delta}$
and $\intarcp{a_\delta}{d_\delta}{c_\delta}{b_\delta}$, respectively, and we call them
internal arc patterns.

We consider the quantity
\begin{equation}
M_t^\delta = \P_\delta ( \,\intarcp{a_\delta}{d_\delta}{c_\delta}{b_\delta}\, |\, \F_t) 
\end{equation}
which we call an observable. 
Here $\F_t$ is the $\sigma$-algebra generated by $\gamma(s)$, $s \in [0,t]$. 
Since $M_t^\delta$ is a conditional expected value 
of a random variable with respect to
$\F_t$, the process $(M_t^\delta)_{t \geq 0}$ 
is a martingale with respect to the filtration $(\F_t)_{t \geq 0}$ and
the probability measure $\P_\delta$.

\subsection{Scaling limit of the observable}

In \cite{Kemppainen:2015vu}, it was shown that the observables $M_t^\delta$ 
converge to a scaling limit $M_t$.
It has an explicit formula
\begin{equation}\label{eq: m formula}
M_t  =  \sqrt{1 + \frac{Y_t}{X_t}} - \sqrt{\frac{Y_t}{X_t}} .
\end{equation}
The mode of convergence is given by the following result:

\begin{proposition}\label{prop: unif conv of martingales}
For each $\eps>0$ and $T>0$, there exists an event $E$ and $\delta_0 >0$ such that the following holds.
If $\delta \leq \delta_0$, then $\P_\delta(E) > 1- \eps$ and
\begin{equation*}
\sup_E \sup_{t \in [0,T]} |M_t^{\delta} - M_t| \leq \eps .
\end{equation*}
\end{proposition}

It follows that also $M_t$ is a martingale. Namely, let $s<t$ and let $f$ be any continuous, bounded random variable
which is measurable with respect to $\F_s$. 
Then $\E_\delta (M_t^\delta f) = \E_\delta (M_s^\delta f)$ by the martingale
property of the discrete observable. By the triangle inequality
\begin{align*}
|\E(M_t f) - \E(M_s f)| \leq & \; |\E(M_t f) - \E_\delta (M_t f)| + |\E(M_s f)-\E_\delta (M_s f)| \\
  &+ |\E_\delta ((M_t-M_t^\delta) f)| + |\E_\delta ((M_s-M_s^\delta) f)| .
\end{align*}
First and second term tend to zero as $\delta \to 0$ by the weak convergence of probability measures.
The third and fourth term also tend to zero by Proposition~\ref{prop: unif conv of martingales}, since
$|\E_\delta ((M_t-M_t^\delta) f)| 
  \leq 2 \,\E_\delta ( \ind_{E^c} |f| ) 
       + \sup_E \sup_{t \in [0,T]} \; |M_t^{\delta} - M_t| \, \E_\delta(|f|)$.

\subsection{Weighting by a martingale}

We weight the probability measure by the martingale $M_t/M_0$ 
(the process is stopped upon the martingale hitting 
$0$ or $1$, i.e. when $X_t$ or $Y_t$ hit zero).

Denote the event $\intarcp{a_\delta}{d_\delta}{c_\delta}{b_\delta}$ by $A$.
Then by properties of conditional expected values
\begin{equation*}
\E( f M_t)=\E( f \, \E(\ind_A\,|\,\F_t))=\E( \E(f\,\ind_A\,|\,\F_t))  = E(f \,\ind_A)
\end{equation*}
for any $\F_t$-measurable bounded random variable $f$. Thus the probability
measure $\P$ weighted by $M_t/M_0$ can be interpreted as to be conditioned
by the event ${\intarcp{a_\delta}{d_\delta}{c_\delta}{b_\delta}}$
and thus equals to $\P^+$.

\subsubsection{Girsanov's theorem}\label{sssec: girsanov}

Suppose that $N_t$ is a martingale such that
\begin{equation*}
M_t = \exp \left( N_t - \frac{1}{2} \langle N \rangle_t  \right)
\end{equation*}
Then by It\^o's lemma, $M_t$ and $N_t$ satisfy the identity
\begin{equation*}
N_t = N_0 + \int_0^t \frac{\de M_s}{M_s}
\end{equation*}
which can be used for defining $N_t$ for any positive martingale $M_t$.

Under the  probability measure weighted by the martingale $M_t/M_0$, it holds that
the process
\begin{equation}\label{eq: def new BM}
B_t - \langle B, N \rangle_t 
\end{equation}
is a standard Brownian motion by Girsanov's theorem (see for instance \cite{Durrett:1996wh}, Section~2.12).
Thus if we have a Loewner evolution whose driving process is
\begin{equation*}
U_t = U_0 + \sqrt{\kappa} B_t + D_t
\end{equation*}
where $D_t$ is the drift of $W_t$ in the sense that $D_t$ is a bounded variation process,
then the driving process can be written as
\begin{equation*}
U_t = U_0 + \sqrt{\kappa} \hat B_t + D_t + \Delta_t
\end{equation*}
where $\hat B_t$ is a standard Brownian motion under the weighted probability measure. Here 
\begin{equation*}
\Delta_t = \sqrt{\kappa} \langle B, N \rangle_t 
\end{equation*}
by \eqref{eq: def new BM}.

\subsection{The driving process conditioned on the internal arc configuration}

Remember that by results in \cite{Kemppainen:2015vu},
\begin{equation*}
M_t = M_0 + \int_0^t \frac{1}{2 \sqrt{3}} \frac{(1-M_s^2)^3}{Y_s\,M_s (M_s^2+1)} \de B_s
\end{equation*}
and
\begin{equation*} 
\de U_t = -\frac{4}{\sqrt{3}} \de B_t + 
  \left( \frac{2}{X_t} - 
  \frac{1}{3} \frac{(3M_t^4+2M_t^2+1)(1-M_t^2)^2}{Y_t M_t^2 (M_t^2+1)^2} \right) \de t .
\end{equation*}
Consequently by the considerations of Section~\ref{sssec: girsanov},
for a process $(\hat B_t)$ which is a Brownian motion under the measure
 $\P^+$ (the one weighted by $(M_t /M_0)$), it holds that
\begin{align*} 
\de U_t &= -\frac{4}{\sqrt{3}} \de \hat B_t + 
  \left( \frac{2}{X_t}  
  -\frac{1}{3} \frac{(3M_t^4+2M_t^2+1)(1-M_t^2)^2}{Y_t M_t^2 (M_t^2+1)^2} 
  -\frac{2}{3} \frac{(1-M_s^2)^3}{Y_s\,M_s^2 (M_s^2+1)}
  \right) \de t  \nonumber \\
  & = -\frac{4}{\sqrt{3}} \de \hat B_t + 
  \left( \frac{2}{X_t}  
  -\frac{1}{3} \frac{(M_t^4+2M_t^2+3)(1-M_t^2)^2}{Y_t M_t^2 (M_t^2+1)^2} 
  \right) \de t   .
\end{align*}
The rightmost term on the first line is $ \sqrt{\kappa} \,\de\langle B, N \rangle_t $.
By plugging in the expression~\eqref{eq: m formula} 
gives after some algebra
\begin{align*} 
\de U_t & = -\frac{4}{\sqrt{3}} \de \hat B_t + \left( \frac{2}{X_t} - \frac{2}{X_t+Y_t} 
   - \frac{4}{3} \frac{Y_t\left( 2 + \sqrt{1+\frac{X_t}{Y_t}}\right)}{X_t(X_t+Y_t)}
  \right) \de t 
\end{align*}
which is equivalent to \eqref{eq: hsle driving}. Thus it follows that $\P^+$
is the law of a hypergeometric \sle{\frac{16}{3}}.

\subsection{Joint law of the pair of interfaces in the arc configuration}
  \label{ssec: joint law}

The scaling limit of the joint law of the interfaces from $a_\delta$ to $d_\delta$
and $c_\delta$ to $b_\delta$ under the probability measure conditioned on the
event $\intarcp{a_\delta}{d_\delta}{c_\delta}{b_\delta}$ can be characterized in the 
following way. 

Consider the scaling limit of the pair of interfaces in the conditioned arc configuration
after conformal transformation to the upper half-plane and
let the curves be $\gamma_1$ and $\gamma_2$ such that
$\gamma_1$ and $\gamma_2$ start at $U_0$ and $W_0$, respectively.
Parametrize the curves $\gamma_1$ and $\gamma_2$ in some way.
For instance, use the half-plane capacity seen from $\infty$ or $V_0$
as parametrization for $\gamma_1$ and $\gamma_2$, respectively.
Then define $\F_{s,t}$ to be the $\sigma$-algebra generated
by $\gamma_1(q)$, $q \in [0,s]$, and $\gamma_2(r)$, $r \in [0,t]$.

By the same argument that says that the marginal law of $\gamma_1$ is
the hSLE, we see that conditionally on $\F_{s,t}$, the marginal law of $\gamma_1$ is
the hSLE. Degenerate versions of these statements give that (i) 
the pair $(\gamma_1,\gamma_2)$ can be sampled by sampling first $\gamma_j$, $j=1$ or $2$, as hSLE
in $\half$ and then sampling $\gamma_{3-j}$ in $H$,
where $H$ is the component of $\gamma_{3-j}(0) + \ii \,( 0+)$ in $\half \setminus \gamma_j (0,\infty)$,
as an independent chordal SLE
and $(ii)$ that a similar conditional version holds (i.e. conditional on $\F_{s,t}$, the pair can be
sampled as an hSLE and an independent chordal SLE).

\section{Comparison to a similar result on percolation}

Let us compare the previous case of FK Ising model to that of 
the critical site percolation model on triangular lattice.

Consider the site percolation model on the triangular lattice 
$${\delta L_\textnormal{tri}} = {\{\delta( j +k e^{\ii \pi /3})\,:\, j,k \in \Z\}}.$$
It was shown in \cite{Smirnov:2001hw}, that the interface of this model in the chordal setup
converges to \sle{6}. 

Note that the existence
of an open percolation crossing from $[a_\delta b_\delta]$ to $[c_\delta d_\delta]$ in $\Omega_\delta$
is exactly the event of an internal arc pattern
$\intarcp{a_\delta}{d_\delta}{c_\delta}{b_\delta}$ of interfaces.
A central result in \cite{Smirnov:2001hw} 
is that the probability of such a crossing event is given by Cardy's formula 
\begin{equation*}
\lim_{\delta \to 0} \P_\delta(\intarcp{a_\delta}{d_\delta}{c_\delta}{b_\delta}) 
 = C \, \ffrac{X_0}{X_0+Y_0}^{\frac{1}{3}} \, 
   \vphantom{F}_2 F_1 \left(\frac{1}{3}, \frac{2}{3}, \frac{4}{3}; \frac{X_0}{X_0+Y_0}\right)
\end{equation*}
holds, where $X_0 = V_0-U_0$ and $Y_0 = W_0-V_0$ with the notation used above and $C$ is a constant, 
whose exact value we don't need below.

It follows then that
\begin{equation*}
M_t = \ffrac{X_t}{X_t+Y_t}^{\frac{1}{3}} 
      \, \vphantom{F}_2 F_1 \left(\frac{1}{3}, \frac{2}{3}, \frac{4}{3}; \frac{X_t}{X_t+Y_t}\right)
\end{equation*}
is a martingale for the scaling limit for $t \leq \tau$ where $\tau$ is the time when the quarilateral
degenerates (the interface hits $[bc]$ or $[cd]$).
Since the interface (exploration process from $a$ to $d$) converges to the chordal \sle{6},
\begin{equation*}
\de U_t = \sqrt{6} \, \de B_t , \qquad
\de X_t = -\sqrt{6} \, \de B_t+ \frac{2}{X_t}, \qquad
\partial_t Y_t = \frac{2}{X_t+Y_t} - \frac{2}{X_t}  .
\end{equation*}
Thus it follows that if $\de N_t = \frac{\de M_t}{M_t}$
\begin{equation*}
\sqrt{6} \, \de\langle B,N \rangle_t = -\frac{2 \ffrac{Y_t}{X_t+Y_t}^\frac{1}{3}}{
 X_t \,\vphantom{F}_2 F_1 \left(\frac{1}{3}, \frac{2}{3}, \frac{4}{3}; \frac{X_t}{X_t+Y_t}\right)} \de t
\end{equation*}
Thus for the process $\hat{B}_t=B_t - \langle B, N \rangle_t $ which is a Brownian motion
under the probability measure weighted by the martingale $M_t/M_0$, the driving process $U_t$ satisfies 
\begin{align*} 
\de U_t & = \sqrt{6} \de \hat B_t -\frac{2 \ffrac{Y_t}{X_t+Y_t}^\frac{1}{3}}{
 X_t \,\vphantom{F}_2 F_1 \left(\frac{1}{3}, \frac{2}{3}, \frac{4}{3}; \frac{X_t}{X_t+Y_t}\right)} \de t
\end{align*}
which shows that the Loewner evolution is the hypergeometric \sle{6}.

\section*{Acknowledgements}

AK was supported by the Academy of Finland.
SS was supported by the ERC
AG COMPASP, the NCCR SwissMAP, the Swiss NSF, and the Russian Science Foundation.

\nocite{*}

\bibliographystyle{habbrv}
\bibliography{fki-hsle.bib}

\end{document}